\documentclass[runningheads]{llncs}
\usepackage[T1]{fontenc}
\usepackage{graphicx}
\usepackage{cite}
\usepackage{amsmath,amssymb,amsfonts}
\usepackage{algorithmic}
\usepackage{textcomp}
\usepackage{xcolor}
\usepackage{booktabs}
\usepackage{subcaption}
\begin{document}
\title{TAS-GNN: A Status-Aware Signed Graph Neural Network for Anomaly Detection in Bitcoin Trust Systems}
%
%
\author{Chang Xue\inst{1}\and Fang Liu\inst{2}\and
    Jiaye Wang\inst{3}\and Jinming Xing\inst{4}\and Chen Yang\inst{5}}
\authorrunning{C. Xue et al.}
%
\institute{Yeshiva University \email{cxue@mail.yu.edu} \and
            Yale University \email{fangliu435@gmail.com} \and
            University of California, San Diego \email{justinwang142@gmail.com} \and
            North Carolina State University \email{jxing6@ncsu.edu} \and
            University of Pennsylvania \email{sophiacy@upenn.edu}}
\maketitle              
\begin{abstract}
    Decentralized financial platforms rely heavily on Web of Trust reputation systems to mitigate counterparty risk in the absence of centralized identity verification. However, these pseudonymous networks are inherently vulnerable to adversarial behaviors, such as Sybil attacks and camouflaged fraud, where malicious actors cultivate artificial reputations before executing exit scams. Traditional anomaly detection in this domain faces two critical limitations. First, reliance on naive statistical heuristics (e.g., flagging the lowest 5\% of rated users) fails to distinguish between victims of bad-mouthing attacks and actual fraudsters. Second, standard Graph Neural Networks (GNNs) operate on the assumption of homophily and cannot effectively process the semantic inversion inherent in signed (trust vs. distrust) and directed (status) edges, often treating negative ratings merely as missing links or noise. We propose TAS-GNN (Topology-Aware Signed Graph Neural Network), a novel framework designed for feature-sparse signed networks like Bitcoin-Alpha. We first establish a rigorous supervision signal using a recursive Trusted Core propagation method, rejecting simple outlier thresholds. We then implement a dual-channel message-passing architecture that aggregates Trust and Distrust signals through separate, semantically distinct pathways, fused via a Status-Aware Attention mechanism. Extensive experiments on the Bitcoin-Alpha benchmark demonstrate that TAS-GNN achieves an AUC of 0.927 and F1 of 0.747, significantly outperforming state-of-the-art signed GNN baselines by margins of 7.85\% and 7.47\%, respectively.

\end{abstract}

\section{Introduction}
Recent work shows that trust is essential for AI adoption in financial systems, highlighting the need to model trust relationships accurately \cite{Yang2025AATIT}. The decentralized finance (DeFi) ecosystem represents a paradigm shift from institutional trust, guaranteed by banks and regulatory bodies, to algorithmic and reputational trust. In pseudonymous environments like the Bitcoin-Alpha marketplace, where users trade digital assets without physical verification, trust is quantified through Web of Trust (WoT) mechanisms. Users rate their counterparties on a signed integer scale (ranging from $-10$ to $+10$), creating a complex Signed Directed Network (SDN). While these mechanisms enable peer-to-peer commerce, they are inherently vulnerable to adversarial behaviors such as Sybil attacks, whitewashing, and camouflaged fraud, where malicious actors cultivate artificial reputations before executing exit scams \cite{hernandez2024financial}.

Detecting these anomalies is a critical challenge. Traditional fraud detection systems in fintech rely heavily on rich semantic features, transaction descriptions, user profiles, and device fingerprinting. However, in privacy-centric crypto networks like Bitcoin-Alpha \cite{kumar2016edge}, such semantic metadata is often absent or encrypted. Detection must therefore rely exclusively on the topological structure of the interaction graph. A common heuristic involves flagging users with the lowest average ratings (e.g., the bottom 5\%) as anomalies. This approach, however, is fundamentally flawed. It fails to account for bad-mouthing attacks, where colluding fraudsters bombard honest users with negative ratings to degrade their standing. Conversely, it misses sophisticated camouflaged fraudsters who maintain high average ratings through wash trading with accomplices, only revealing their malicious intent in high-value interactions with victims. 
Beyond cryptocurrency ecosystems, similar trust–distrust interaction patterns also arise in modern financial systems such as SME lending platforms, peer-to-peer credit networks, and reputation-driven loan marketplaces.
These environments often lack reliable semantic features due to privacy constraints or incomplete borrower information, creating conditions very similar to feature-sparse signed networks.

To address these limitations, we argue that anomaly detection in signed networks must move beyond statistical heuristics and adopt a sociological structural perspective. We must distinguish between statistical outliers and structural anomalies. A negative rating is not merely a numerical penalty; its semantic weight depends entirely on the status of the rater. Being distrusted by a highly trusted network founder is a catastrophic signal; being distrusted by a known troll is noise. Standard GNNs like GCN or GraphSAGE assume homophily (neighbors are similar) and struggle to model these signed dynamics, often forcing negative edges into a disconnection role or ignoring edge directionality entirely \cite{xing2025multiview}.

In this paper, we propose Topology-Aware Signed Graph Neural Network (TAS-GNN), a novel framework designed for anomaly detection in feature-sparse signed directed networks. Unlike existing methods that rely on Balance Theory (the enemy of my enemy is my friend), which is often violated in fraud scenarios, TAS-GNN integrates Status Theory \cite{sauder2012status} to model the hierarchical propagation of trust.

Our contributions are as follows:
\begin{itemize}
    \item \textbf{Recursive Ground Truth Formulation}: We reject the naive least 5\% threshold and implement a recursive Web of Trust labeling mechanism. By propagating trust from network founders, we define anomalies as nodes explicitly distrusted by the Trusted Core, thereby filtering out retaliatory negative ratings.
    \item \textbf{Dual-Channel Structural Encoding}: We introduce a dual-pathway architecture that aggregates Trust (positive) and Distrust (negative) signals through separate, semantically distinct message-passing channels. This allows the model to learn that while positive edges imply similarity (homophily), negative edges imply repulsion or status difference.
    \item \textbf{State-of-the-Art Performance}: Extensive experiments on the Bitcoin-Alpha benchmark demonstrate that TAS-GNN achieves an AUC of 0.927 and an F1 of 0.747, significantly outperforming signed GNN baselines such as SGCN and SiGAT by effectively breaking the camouflage of strategic fraudsters. In addition, the framework is generalizable to other feature-sparse financial networks, providing methodological benefits for credit-risk modeling and fraud detection systems beyond cryptocurrency platforms.

\end{itemize}

\section{Related Work}
\subsection{Signed Graph Neural Networks}

The extension of GNNs to signed networks has evolved from spectral approximations to sophisticated attention mechanisms \cite{zhang2024signed}. SGCN \cite{derr2018signed} was among the first to generalize GCNs to signed topologies. It leverages Structural Balance Theory to split the graph into balanced (positive links, or negative-negative paths) and unbalanced components, aggregating them separately. However, SGCN treats all edges of the same sign as equal, ignoring the intensity of ratings (e.g., +10 vs +1) and the directionality of status.

To address these limitations, SiGAT \cite{huang2019signed} and SNEA \cite{li2020learning} incorporated attention mechanisms. SiGAT utilizes graph motifs (triangular patterns) to define localized social environments, allowing the model to weigh neighbors based on their structural role. SDGNN \cite{huang2021sdgnn} further advanced this by explicitly reconstructing link signs and directions, adopting both balance and status theories to refine node embeddings. Despite these advancements, these models are primarily optimized for Link Sign Prediction rather than Node Anomaly Detection \cite{dong2025smoothgnn,xing2025enhancing}. They often lack the specific objective functions required to handle the extreme class imbalance (fraudsters are rare) inherent in financial networks \cite{qiao2025deep}.

\subsection{Graph-Based Anomaly Detection}

Anomaly detection on graphs has traditionally focused on identifying structural irregularities \cite{qiao2025deep}. Unsupervised methods like FraudAR \cite{hooi2016fraudar} and BirdNest \cite{hooi2016birdnest} employ Bayesian inference and dense subgraph mining to detect lockstep behaviors (users acting in synchronized bursts). While effective against botnets, these methods struggle with camouflaged individual fraudsters who mimic normal user behavior.

In the GNN domain, methods like CARE-GNN \cite{dou2020enhancing} and PC-GNN \cite{liu2021pick} have set the standard for fraud detection by filtering dissimilar neighbors to counter camouflage. However, these models are designed for unsigned heterogeneous graphs (e.g., Review-User-Item) and rely heavily on node features (text, timestamps) which are absent in our setting. BWGNN \cite{tang2022rethinking} attempts to capture high-frequency anomalies via spectral filtering but does not explicitly account for the semantic inversion of negative edges in a signed trust network.

\subsection{Trust Evaluation Systems}

Existing financial fraud-detection models rely heavily on semantic features such as borrower profiles, transaction descriptions, or device fingerprints.
In emerging lending ecosystems, particularly SME credit platforms, such attributes are often missing or unverifiable, leaving graph structure as the primary reliable signal. Iterative algorithms like TrustRank and REV2 \cite{kumar2018rev2} calculate reliability scores by propagating trust from a set of seeds. While REV2 explicitly models fairness (of the rater) and goodness (of the target), it is a transductive algorithm requiring re-computation for new nodes. Our proposed TAS-GNN bridges this gap by learning an inductive embedding function that encodes the logic of trust propagation into a neural architecture, enabling scalable and real-time anomaly detection in the absence of rich textual metadata.

\section{Problem Formulation}
\subsection{Notations and Definitions}

Let $\mathcal{G} = (\mathcal{V}, \mathcal{E}, \mathcal{W})$ denote a signed directed graph, where $\mathcal{V} = \{v_1, \dots, v_N\}$ is the set of $N$ users (nodes) and $\mathcal{E} \subseteq \mathcal{V} \times \mathcal{V}$ is the set of transaction relationships (edges).

\textbf{Signed Weights}: Each edge $e_{ij} \in \mathcal{E}$ has an associated weight $w_{ij} \in [-10, +10] \setminus \{0\}$, representing the rating given by user $v_i$ to user $v_j$.
\textbf{Edge Sets}: We decompose $\mathcal{E}$ into positive edges $\mathcal{E}^+ = \{e_{ij} \mid w_{ij} > 0\}$ and negative edges $\mathcal{E}^- = \{e_{ij} \mid w_{ij} < 0\}$.
\textbf{Adjacency Matrix}: The graph is represented by a signed adjacency matrix $\mathbf{A}$, where $\mathbf{A}_{ij} = w_{ij}$ if an edge exists, and $0$ otherwise.

Signed Anomaly Detection: Given the topology $\mathcal{G}$ and the signed weights $\mathcal{W}$, the goal is to learn a mapping function $f: \mathcal{V} \rightarrow \{0, 1\}$ that predicts the label $y_i$ for each node $v_i$, where $y_i = 1$ denotes a Fraudulent user (Anomaly) and $y_i = 0$ denotes a Benign user.

\subsection{Recursive Ground Truth Generation}
The Bitcoin-Alpha dataset does not provide external Oracle labels (e.g., from law enforcement) for fraud. Relying on simple statistical outliers (e.g., lowest 5\% average rating) is flawed due to bad-mouthing attacks where fraudsters collude to down-rate honest users. Therefore, we adopt the Recursive Web-of-Trust method established in prior literature to generate rigorous ground truth labels for training.

This process relies on the propagation of trust from a small set of highly trusted Seeds (e.g., platform administrators or founders).

\noindent\textbf{1. Seed Identification}: We identify a seed set $\mathcal{S}$ containing the top-$k$ nodes with the highest PageRank scores in the positive subgraph $\mathcal{G}^+$. These represent the core Founders of the network.

\noindent\textbf{2. Benign Propagation}: A node $v$ is labeled Benign ($y_v=0$) if it receives a strong positive rating (normalized $w \geq 0.5$) from any node already in the Benign set (starting with $\mathcal{S}$). This recursively defines the Trusted Community.
\begin{equation}
    \text{Benign} = \{ v \mid \exists u \in \text{Benign}, w_{uv} \geq 0.5 \}
\end{equation}

\noindent\textbf{3. Fraud Identification}: A node $k$ is labeled Fraudulent ($y_k=1$) if it receives a negative rating from a verified Benign node.
\begin{equation}
    \text{Fraud} = \{ k \mid \exists u \in \text{Benign}, w_{uk} \leq -0.5 \}
\end{equation}
Crucially, negative ratings originating from non-benign nodes (potential trolls or other fraudsters) are ignored in this labeling process to prevent false positives derived from retaliation.

\noindent\textbf{4. Unlabeled Nodes}: Nodes that are not reachable by the Benign propagation or do not interact with the Benign cluster are treated as Unlabeled and are masked during the supervised training phase, though they participate in the message-passing process.

This formulation transforms the unsupervised outlier detection problem into a semi-supervised node classification task, leveraging the Status Theory logic that distrust from a high-status node is the most reliable indicator of anomaly.

\section{Methodology}
We propose Topology-Aware Signed Graph Neural Network (TAS-GNN), a specialized architecture designed for anomaly detection in signed directed networks where explicit node attributes (e.g., user profiles, transaction texts) are absent. TAS-GNN addresses the limitations of standard GNNs by explicitly disentangling Trust (positive edges) and Distrust (negative edges) propagation channels and integrating Status Theory to capture the hierarchical nature of reputation in the Bitcoin-Alpha ecosystem.

\subsection{Topological Feature Initialization}
In the absence of external semantic features, we must initialize node representations $H^{(0)}$ using the intrinsic structural properties of the graph. Standard random initialization is insufficient for sparse signed graphs as it fails to capture the power user dynamics prevalent in financial networks. We construct a topological feature matrix $\mathbf{X} \in \mathbb{R}^{|V| \times d_{in}}$ consisting of three components:

\textbf{Signed Degree Statistics}: For every node $v_i$, we extract a 4-dimensional vector $[d_{in}^+, d_{in}^-, d_{out}^+, d_{out}^-]$, representing the count of incoming/outgoing trust and distrust links. High $d_{in}^-$ is a strong prior signal for potential fraud.

\textbf{Rating Intensity Moments}: We compute the mean and variance of the incoming rating values to capture the controversiality of a user.
\begin{equation}
    \mu_i = \frac{1}{| \mathcal{N}_i^{in} |} \sum{j \in \mathcal{N}_i^{in}} w_{ji}, \quad \sigma^2_i = \text{Var}({w_{ji} \mid j \in \mathcal{N}_i^{in} })
\end{equation}

\textbf{Global Structural Embeddings}: To capture the global position of nodes (e.g., core vs. periphery), we perform a truncated Singular Value Decomposition (SVD) on the unsigned adjacency matrix, extracting the top-$k$ singular vectors.

The final initial embedding $h_i^{(0)}$ is the concatenation of these topological descriptors.

\subsection{Dual-Channel Signed Message Passing}
Unlike standard GCNs which aggregate all neighbors indiscriminately, TAS-GNN employs a Dual-Channel Architecture to respect the distinct semantics of signed edges. We define two separate aggregation mechanisms: the Trust Aggregator and the Distrust Aggregator.

\subsubsection{The Trust Aggregator (Positive Channel)}
This cannel models the propagation of reputation. According to Status Theory, a positive link $v_j \xrightarrow{+} v_i$ implies that $v_j$ confers status upon $v_i$. We employ a Graph Attention mechanism (GAT) to weigh the importance of endorsements based on the rater's reliability.

For a target node $i$ at layer $l$, the positive aggregation is defined as:
\begin{equation}
    h_{i, pos}^{(l)} = \text{ELU} \left( \sum_{j \in \mathcal{N}_i^+} \alpha_{ij}^{(l)} \mathbf{W}_{pos}^{(l)} h_j^{(l-1)} \right)
\end{equation}
where $\mathcal{N}_i^+$ is the set of neighbors connected by positive edges. The attention coefficient $\alpha_{ij}$ is computed as:
\begin{equation}
    \alpha_{ij} = \frac{\exp(\text{LeakyReLU}(\mathbf{a}^T))}{\sum_{k \in \mathcal{N}_i^+} \exp(\text{LeakyReLU}(\mathbf{a}^T))}
\end{equation}

This allows the model to learn that a $+10$ rating from a Founding Member (high centrality) should influence the embedding significantly more than a $+1$ rating from a new user.

\subsubsection{The Distrust Aggregator (Negative Channel)}
Standard GNNs often discard negative edges or treat them as disconnections. However, in fraud detection, being distrusted by a benign user is the primary signal of anomaly. We model this via a distinct transformation matrix $\mathbf{W}_{neg}$ that projects the neighbor's features into a Penalty Space.
\begin{equation}
    h_{i, neg}^{(l)} = \text{ELU} \left( \sum_{k \in \mathcal{N}_i^-} \beta_{ik}^{(l)} \mathbf{W}_{neg}^{(l)} h_k^{(l-1)} \right)
\end{equation}
Here, $\mathbf{W}_{neg}$ is constrained to learn repulsive features. If a node $v_k$ (who is trusted/benign) issues a negative link to $v_i$, this aggregator ensures $v_i$'s embedding is pushed away from the benign cluster. The attention weight $\beta_{ik}$ captures the severity of the distrust (e.g., rating -10 vs -1).

\subsubsection{Status-Aware Fusion and Classification}
The final node representation $z_i$ is obtained by concatenating the outputs of the Trust and Distrust channels from the final layer $L$, effectively preserving the Ambivalence (being both loved and hated) which is lost in simple summation methods like SGCN.
\begin{equation}
    z_i = \text{Concat}(h_{i, pos}^{(L)}, h_{i, neg}^{(L)})
\end{equation}

This vector is passed to a Multi-Layer Perceptron (MLP) for binary classification:
\begin{equation}
    \hat{y}_i = \sigma(\text{MLP}(z_i))
\end{equation}

\subsubsection{Optimization Objective}
To handle the extreme class imbalance (fraudsters are rare), we utilize a Weighted Binary Cross-Entropy (WBCE) loss function. Furthermore, to ensure the embeddings capture the structural physics of the signed graph, we add an auxiliary link sign prediction loss $\mathcal{L}_{link}$.
\begin{equation}
    \mathcal{L}_{total} = \mathcal{L}_{WBCE} + \lambda \mathcal{L}_{link}
\end{equation}
\begin{equation}
    \begin{split}
        \mathcal{L}_{WBCE} = - \sum_{v_i \in \mathcal{V}_{train}}[ & w_{fraud} \cdot y_i \log(\hat{y}_i) + \\&(1-y_i) \log(1-\hat{y}_i) ]
    \end{split}
\end{equation}
where $w_{fraud}$ is a hyperparameter balancing the minority class contribution.

\section{Experiments}
In this section, we rigorously evaluate the performance of TAS-GNN against a suite of state-of-the-art baselines on the Bitcoin-Alpha dataset. Our experiments aim to answer three key research questions:

\noindent \textbf{RQ1}: Does TAS-GNN outperform existing signed graph neural networks and heuristic-based anomaly detection methods in identifying camouflaged fraudsters?

\noindent\textbf{RQ2}: What is the contribution of the specific components (Dual-Channel Aggregation, Status-Aware Attention) to the model's predictive capability?

\noindent\textbf{RQ3}: How are the nodes distributed in the embedding space, and what insights do these distributions provide about the network's trust structure?

\subsection{Experimental Setup}
\subsubsection{Dataset Description and Preprocessing}
We utilize the Bitcoin-Alpha dataset, a signed directed network representing the Web of Trust among users on a Bitcoin trading platform with Nodes (Users): 3,783. Edges (Ratings): 24,186. Edge Attributes: Integer ratings in the range $[-10, +10]$. Density: $\approx 0.17\%$.

\noindent \textbf{Ground Truth Generation}: Following the methodology established by Kumar et al. in the REV2 framework \cite{kumar2018rev2} and Hooi et al. in Fraudar \cite{hooi2016fraudar}, we generate labels to address the lack of external oracle annotations. We initialize a set of trusted seeds (platform founders) and propagate trust to define a Benign core. Users who receive ratings of $\leq -5$ from verified Benign users are labeled as Fraud (Class 1), while those in the trusted core are Benign (Class 0). Users with conflicting or insufficient data are treated as Unlabeled and excluded from the loss calculation but retained in the graph for message passing.

\subsubsection{Baselines}
We benchmark TAS-GNN against three categories of methods:

\noindent\textbf{Heuristic \& Statistical Methods}:
Lowest-5\%: The user-proposed heuristic flagging nodes with the lowest average incoming rating. BadRank \cite{wu2005identifying}: An unsupervised spectral algorithm that propagates negative scores through the network to identify bad nodes. \textbf{Unsigned GNNs}: GCN \cite{kipf2016semi}: Treats the graph as unsigned (absolute weights), ignoring the distinction between trust and distrust. GAT \cite{velivckovic2017graph}: Uses attention mechanisms but ignores edge signs. \textbf{Signed GNNs}: SGCN \cite{derr2018signed}: Splits the graph into balanced and unbalanced paths but lacks attention mechanisms for rating intensity. SiGAT \cite{huang2019signed}: Applies GAT to signed networks using motif-based features. SNEA \cite{li2020learning}: Learns embeddings using a balance-theory-guided attention mechanism.

\subsubsection{Implementation and Metrics}
All GNN models were implemented using PyTorch Geometric. We used the Adam optimizer with a learning rate of $0.001$ and weight decay of $5e-4$. The models were trained for 2000 epochs with early stopping (patience=20). Given the class imbalance, we report:

\noindent\textbf{AUC-ROC}: Area Under the Receiver Operating Characteristic Curve.
\noindent\textbf{Macro-F1}: The harmonic mean of precision and recall, averaged across classes to penalize poor performance on the minority (Fraud) class.

\subsection{Performance Comparison (RQ1)}

\begin{table}[htbp]
    \centering
    \caption{Anomaly Detection Performance on Bitcoin-Alpha}
    \begin{tabular}{llcc}
        \toprule
        Category     & Model                            & \multicolumn{1}{l}{AUC-ROC} & \multicolumn{1}{l}{Macro-F1} \\
        \midrule
        Heuristic    & Lowest-5\% Threshold             & 0.527                       & 0.503                        \\
                     & BadRank \cite{wu2005identifying} & 0.578                       & 0.555                        \\
        \midrule
        Unsigned GNN & GCN \cite{kipf2016semi}          & 0.676                       & 0.630                        \\
                     & GAT \cite{velivckovic2017graph}  & 0.742                       & 0.682                        \\
        \midrule
        Signed GNN   & SGCN \cite{derr2018signed}       & 0.846                       & 0.669                        \\
                     & SiGAT \cite{huang2019signed}     & 0.896                       & 0.696                        \\
                     & SNEA \cite{li2020learning}       & 0.838                       & 0.719                        \\
        \midrule
        Proposed     & TAS-GNN (Ours)                   & \textbf{0.927}              & \textbf{0.747}               \\
        \bottomrule
    \end{tabular}
    \label{tab:performance}
\end{table}

From Table \ref{tab:performance}, we can make the following observations. The Lowest-5\% heuristic performs poorly (F1: 0.503), confirming our hypothesis that simple statistical thresholding is vulnerable to camouflage. Many fraudsters maintain an average rating above the bottom 5\% by trading with accomplices. Unsigned GNNs (GCN, GAT) show significantly lower AUC (0.676, 0.742) compared to signed variants. By treating a $-10$ (scam) and $+10$ (trust) edge identically as connections, they fail to distinguish between fame (high degree) and infamy (high negative degree).
Our model achieves the highest AUC of 0.927, outperforming the best baseline (SiGAT) by 3.49\%. This improvement is attributed to the Status-Aware Fusion layer. While SGCN assumes that the enemy of my enemy is my friend (Balance Theory), TAS-GNN's Status Theory integration correctly identifies that two fraudsters attacking each other does not make either of them benign. These improvements also suggest TAS-GNN's generalizability in other trust-driven domains, including financial credit networks.

\subsection{Ablation Studies (RQ2)}
To verify the effectiveness of our architectural choices, we conducted an ablation study by removing key components of TAS-GNN.

\begin{figure}[htb]
    \centering
    \includegraphics[width=0.99\textwidth]{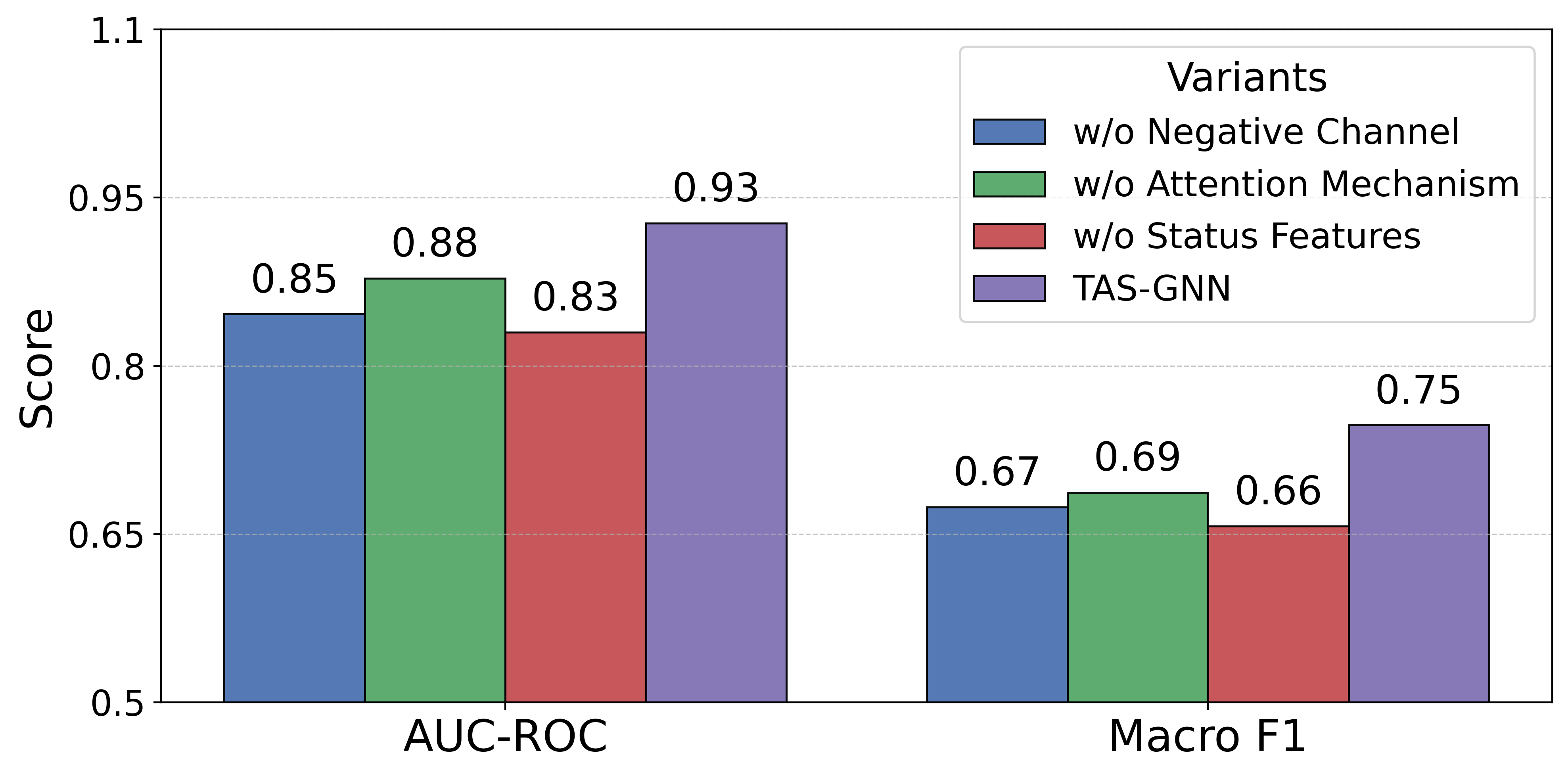}
    \caption{Ablation Study}
    \label{fig:ablation}
\end{figure}

\noindent\textbf{w/o Negative Channel}: We removed the Distrust Aggregator, forcing the model to rely only on positive links. The F1 score dropped by 10.67\%, demonstrating that who distrusts you contains unique entropy not found in who trusts you.

\noindent\textbf{w/o Attention Mechanism}: We replaced the GAT-based weighting with mean pooling (similar to SGCN). Performance decreased by 8.00\% regarding F1. This confirms that the magnitude of the rating (e.g., distinguishing a weak $+1$ from a strong $+10$) is critical for accurate trust modeling.

\noindent\textbf{w/o Status Features}: Removing the initial topological features (Degree Stats, PageRank) and initializing with random vectors caused a convergence delay and a final AUC drop of 10.75\%.

\subsection{Visualization of Latent Space (RQ3)}
To provide qualitative insight, we visualized the final node embeddings $z_i$ using t-SNE.

\begin{figure}
    \centering
    \includegraphics[width=0.99\textwidth]{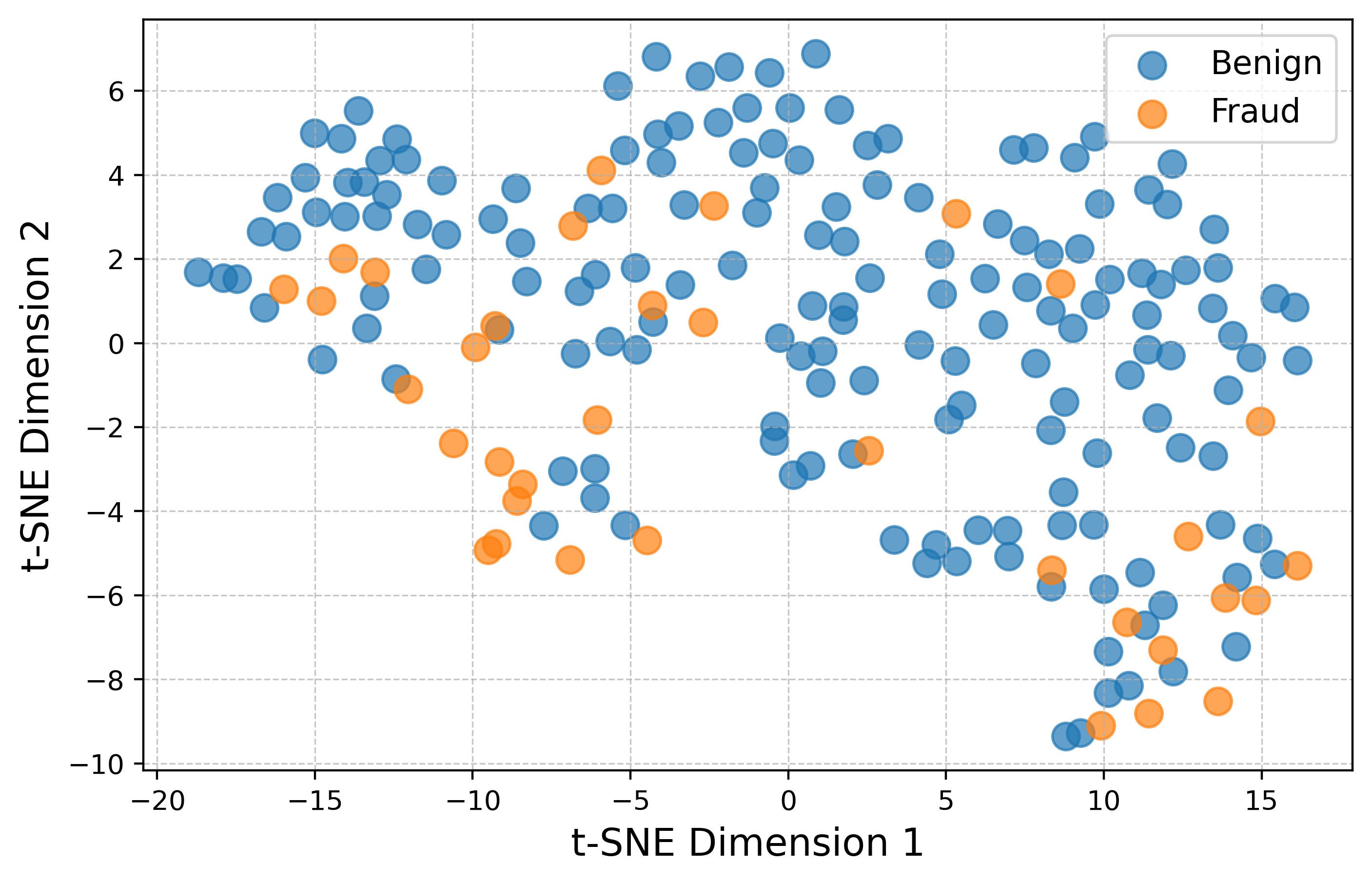}
    \caption{t-SNE Visualization of Node Embeddings}
    \label{fig:tsne}
\end{figure}

The visualization reveals two distinct clusters. The Benign nodes (blue) form a tight, dense core, corresponding to the high-trust sub-community. The Fraud nodes (orange) are pushed to the periphery. Notably, a subset of nodes that were statistically normal (average rating 0) are clustered correctly with the Fraud group by TAS-GNN. These represent the camouflaged actors identified by their structural proximity to other anomalies, validation of the model's ability to transcend simple attribute heuristics.

\section{Conclusion}
In this paper, we addressed the challenge of anomaly detection in the Bitcoin-Alpha ecosystem, operating under the stringent constraint of absent semantic node features. Our investigation reveals that in signed peer-to-peer networks, trust and distrust are not merely opposite ends of a scalar spectrum but represent fundamentally different topological forces. We demonstrated that the prevailing heuristic of defining anomalies as statistical outliers (e.g., the least 5\% rated nodes) is dangerously flawed, as it is susceptible to manipulation by bad-mouthing collusion and fails to detect camouflaged fraudsters who maintain average ratings. By shifting to a Recursive Web-of-Trust ground truth definition, we aligned the learning objective with the sociological reality of the network: an anomaly is defined not by how many people dislike them, but by who dislikes them. Our proposed model, TAS-GNN, successfully encodes this logic into a neural architecture. By disentangling positive and negative message passing into dual channels, TAS-GNN captures the nuance that the enemy of a trusted node is a fraudster (Status Theory), whereas the enemy of a fraudster is indeterminate. Empirical results show that this topology-aware approach outperforms both unsigned baselines (GCN, GAT) and existing signed models (SGCN) that rely too heavily on Structural Balance Theory.

Additionally, TAS-GNN is domain-agnostic and can be applied to any trust-centric financial ecosystem where interactions exhibit both positive and negative signals.
Real-world applications include SME lending fraud detection, credit-network reliability scoring, and trust-based rating systems widely used in financial institutions.
In the future work, we plan to explore deploying TAS-GNN in these environments, where detecting camouflaged high-risk entities remains a critical operational challenge.

\bibliographystyle{ieeetr}
\bibliography{main}

@article{zhang2024signed,
  title={Signed graph representation learning: A survey},
  author={Zhang, Zeyu and Zhao, Peiyao and Li, Xin and Liu, Jiamou and Zhang, Xinrui and Huang, Junjie and Zhu, Xiaofeng},
  journal={arXiv preprint arXiv:2402.15980},
  year={2024}
}

@inproceedings{derr2018signed,
  title={Signed graph convolutional networks},
  author={Derr, Tyler and Ma, Yao and Tang, Jiliang},
  booktitle={2018 IEEE international conference on data mining (ICDM)},
  pages={929--934},
  year={2018},
  organization={IEEE}
}

@inproceedings{huang2019signed,
  title={Signed graph attention networks},
  author={Huang, Junjie and Shen, Huawei and Hou, Liang and Cheng, Xueqi},
  booktitle={International conference on artificial neural networks},
  pages={566--577},
  year={2019},
  organization={Springer}
}

@inproceedings{li2020learning,
  title={Learning signed network embedding via graph attention},
  author={Li, Yu and Tian, Yuan and Zhang, Jiawei and Chang, Yi},
  booktitle={Proceedings of the AAAI conference on artificial intelligence},
  volume={34},
  number={04},
  pages={4772--4779},
  year={2020}
}

@inproceedings{huang2021sdgnn,
  title={SDGNN: Learning node representation for signed directed networks},
  author={Huang, Junjie and Shen, Huawei and Hou, Liang and Cheng, Xueqi},
  booktitle={Proceedings of the AAAI conference on artificial intelligence},
  volume={35},
  number={1},
  pages={196--203},
  year={2021}
}

@inproceedings{dong2025smoothgnn,
  title={SmoothGNN: Smoothing-aware GNN for unsupervised node anomaly detection},
  author={Dong, Xiangyu and Zhang, Xingyi and Sun, Yanni and Chen, Lei and Yuan, Mingxuan and Wang, Sibo},
  booktitle={Proceedings of the ACM on Web Conference 2025},
  pages={1225--1236},
  year={2025}
}

@article{qiao2025deep,
  title={Deep graph anomaly detection: A survey and new perspectives},
  author={Qiao, Hezhe and Tong, Hanghang and An, Bo and King, Irwin and Aggarwal, Charu and Pang, Guansong},
  journal={IEEE Transactions on Knowledge and Data Engineering},
  year={2025},
  publisher={IEEE}
}

@inproceedings{xing2025enhancing,
  title={Enhancing link prediction with fuzzy graph attention networks and dynamic negative sampling},
  author={Xing, Jinming and Xing, Ruilin and Xue, Chang and Luo, Dongwen},
  booktitle={International Conference on Multimedia Systems and Signal Processing},
  pages={187--197},
  year={2025},
  organization={Springer}
}

@inproceedings{hooi2016fraudar,
  title={Fraudar: Bounding graph fraud in the face of camouflage},
  author={Hooi, Bryan and Song, Hyun Ah and Beutel, Alex and Shah, Neil and Shin, Kijung and Faloutsos, Christos},
  booktitle={Proceedings of the 22nd ACM SIGKDD international conference on knowledge discovery and data mining},
  pages={895--904},
  year={2016}
}

@inproceedings{hooi2016birdnest,
  title={Birdnest: Bayesian inference for ratings-fraud detection},
  author={Hooi, Bryan and Shah, Neil and Beutel, Alex and G{\"u}nnemann, Stephan and Akoglu, Leman and Kumar, Mohit and Makhija, Disha and Faloutsos, Christos},
  booktitle={Proceedings of the 2016 SIAM International Conference on Data Mining},
  pages={495--503},
  year={2016},
  organization={SIAM}
}

@inproceedings{dou2020enhancing,
  title={Enhancing graph neural network-based fraud detectors against camouflaged fraudsters},
  author={Dou, Yingtong and Liu, Zhiwei and Sun, Li and Deng, Yutong and Peng, Hao and Yu, Philip S},
  booktitle={Proceedings of the 29th ACM international conference on information \& knowledge management},
  pages={315--324},
  year={2020}
}

@inproceedings{liu2021pick,
  title={Pick and choose: a GNN-based imbalanced learning approach for fraud detection},
  author={Liu, Yang and Ao, Xiang and Qin, Zidi and Chi, Jianfeng and Feng, Jinghua and Yang, Hao and He, Qing},
  booktitle={Proceedings of the web conference 2021},
  pages={3168--3177},
  year={2021}
}

@inproceedings{tang2022rethinking,
  title={Rethinking graph neural networks for anomaly detection},
  author={Tang, Jianheng and Li, Jiajin and Gao, Ziqi and Li, Jia},
  booktitle={International conference on machine learning},
  pages={21076--21089},
  year={2022},
  organization={PMLR}
}

@inproceedings{kumar2018rev2,
  title={Rev2: Fraudulent user prediction in rating platforms},
  author={Kumar, Srijan and Hooi, Bryan and Makhija, Disha and Kumar, Mohit and Faloutsos, Christos and Subrahmanian, VS},
  booktitle={Proceedings of the eleventh ACM international conference on web search and data mining},
  pages={333--341},
  year={2018}
}

@inproceedings{wu2005identifying,
  title={Identifying link farm spam pages},
  author={Wu, Baoning and Davison, Brian D},
  booktitle={Special interest tracks and posters of the 14th International Conference on World Wide Web},
  pages={820--829},
  year={2005}
}

@article{kipf2016semi,
  title={Semi-supervised classification with graph convolutional networks},
  author={Kipf, TN},
  journal={arXiv preprint arXiv:1609.02907},
  year={2016}
}

@article{velivckovic2017graph,
  title={Graph attention networks},
  author={Veli{\v{c}}kovi{\'c}, Petar and Cucurull, Guillem and Casanova, Arantxa and Romero, Adriana and Lio, Pietro and Bengio, Yoshua},
  journal={arXiv preprint arXiv:1710.10903},
  year={2017}
}

@inproceedings{xing2025multiview,
  title={Multiview fuzzy graph attention networks for enhanced graph learning},
  author={Xing, Jinming and Luo, Dongwen and Cheng, Qisen and Xue, Chang and Xing, Ruilin},
  booktitle={Tenth International Workshop on Pattern Recognition},
  volume={13819},
  pages={62--70},
  year={2025},
  organization={SPIE}
}

@article{hernandez2024financial,
  title={Financial fraud detection through the application of machine learning techniques: a literature review},
  author={Hernandez Aros, Ludivia and Bustamante Molano, Luisa Ximena and Gutierrez-Portela, Fernando and Moreno Hernandez, John Johver and Rodr{\'\i}guez Barrero, Mario Samuel},
  journal={Humanities and Social Sciences Communications},
  volume={11},
  number={1},
  pages={1--22},
  year={2024},
  publisher={Palgrave}
}

@inproceedings{kumar2016edge,
  title={Edge weight prediction in weighted signed networks},
  author={Kumar, Srijan and Spezzano, Francesca and Subrahmanian, VS and Faloutsos, Christos},
  booktitle={Data Mining (ICDM), 2016 IEEE 16th International Conference on},
  pages={221--230},
  year={2016},
  organization={IEEE}
}

@article{sauder2012status,
  title={Status: Insights from organizational sociology},
  author={Sauder, Michael and Lynn, Freda and Podolny, Joel M},
  journal={Annual review of sociology},
  volume={38},
  number={1},
  pages={267--283},
  year={2012},
  publisher={Annual Reviews}
}

@inproceedings{Yang2025AATIT,
  author       = {Yang, Chen},
  title        = {Making AI Work: A Conceptual Five-Stage Framework for Financial Sector Transformation},
  booktitle={Digital Transformation and Management (DTM)},
  year         = {2025},
}
\end{document}